\documentclass[conference]{IEEEtran}
\IEEEoverridecommandlockouts
\usepackage{cite}
\usepackage{amsmath,amssymb,amsfonts}
\usepackage{algorithm, algorithmic}
\usepackage{graphicx, subcaption}
\usepackage{balance}
\usepackage{textcomp}
\usepackage{xcolor}

\def\BibTeX{{\rm B\kern-.05em{\sc i\kern-.025em b}\kern-.08em
    T\kern-.1667em\lower.7ex\hbox{E}\kern-.125emX}}
\renewcommand{\baselinestretch}{0.975}
\begin{document}
\bstctlcite{IEEEexample:BSTcontrol}
\title{Indoor Coverage Enhancement for RIS-Assisted Communication Systems: Practical Measurements and Efficient Grouping
}

\author{\IEEEauthorblockN{
Sefa Kayraklık\textsuperscript{$\ast\circ$},
Ibrahim Yildirim\textsuperscript{$\ast\bullet$}, 
Yarkın Gevez\textsuperscript{$\ast$}, 
Ertugrul Basar\textsuperscript{$\ast$}, 
Ali Görçin\textsuperscript{$\circ\diamond$}}
	\IEEEauthorblockA{\textsuperscript{$\ast$}CoreLab, Department of Electrical and Electronics Engineering, Koç University, Sariyer 34450, Istanbul, Turkey \\
	\textsuperscript{$\circ$} Communications and Signal Processing Research (HISAR) Lab., TUBITAK BILGEM, Kocaeli, Turkey\\
\textsuperscript{$\bullet$}Faculty of Electrical and Electronics Engineering, Istanbul Technical University, Sariyer 34469, Istanbul, Turkey.\\
\textsuperscript{$\diamond$} Department of Electronics and Communication Engineering, Yildiz Technical University, Istanbul, Turkey\\
		Email:
		skayraklik21@ku.edu.tr,
		yildirimib@itu.edu.tr, ygevez21@ku.edu.tr,
		ebasar@ku.edu.tr,
		agorcin@yildiz.edu.tr
		\vspace{-3ex}}
}

\maketitle

\begin{abstract}
Reconfigurable intelligent surface (RIS)-empowered communications represent exciting prospects as one of the promising technologies capable of meeting the requirements of the sixth generation networks such as low-latency, reliability, and dense connectivity. However, validation of test cases and real-world experiments of RISs are imperative to their practical viability. To this end, this paper presents a physical demonstration of an RIS-assisted communication system in an indoor environment in order to enhance the coverage by increasing the received signal power. We first analyze the performance of the RIS-assisted system for a set of different locations of the receiver and observe around 10 dB improvement in the received signal power by careful RIS phase adjustments. Then, we employ an efficient codebook design for RIS configurations to adjust the RIS states on the move without feedback channels. We also investigate the impact of an efficient grouping of RIS elements, whose objective is to reduce the training time needed to find the optimal RIS configuration. In our extensive experimental measurements, we demonstrate that with the proposed grouping scheme, training time is reduced from one-half to one-eighth by sacrificing only a few dBs in received signal power. 

\end{abstract}


\section{Introduction}






The fifth generation (5G) wireless communication technology has been actively deployed worldwide since its definition at 3GPP Release 15 \cite{Where5G}.
Though 5G technology is designated to offer upgraded mobile broadband and ultra-reliable low-latency communications, it is still uncertain whether it can accommodate upcoming Internet-of-Everything applications \cite{SimRIS_Mag}. In order to address this issue, both academy and industry have been extensively studying cutting-edge technologies and creating innovative concepts for the sixth generation (6G) wireless communication networks. Nevertheless, the demanding and compelling features that come with these innovative and exciting applications include ultra-reliability, low-latency, extremely high data rates, spectrum optimization, and dense connectivity. In order to meet these demanding requirements of 6G, sophisticated physical (PHY) layer approaches have been put forward. Utilizing efficient PHY layer techniques such as ultra massive multiple-input multiple-output (MIMO) systems, index modulation (IM), TeraHertz (THz) frequency bands (over 100 GHz), and reconfigurable intelligent surfaces (RISs) could be the way out to enable the appealing characteristics of 6G.
Recently, researchers are becoming more and more interested in RIS-aided communication due to its clear promise for improving link capacity, minimizing interference, strengthening secure communication, and extending the coverage \cite{Yildirim_multiRIS,liu2021reconfigurable,SimRIS_TCOM}. RIS-aided communication has therefore begun to be considered a strong candidate to become a possible component of 6G systems alongside other strong candidates such as ultra-massive MIMO and THz communication \cite{RISParadigm6G,RIS_AB_HPC}.

As a consequence of the above-mentioned expected benefits, it is evident that RIS-aided schemes gleam as an attractive pathway for wireless communications. In order to address certain requirements of 6G, researchers have recently studied various attractive RIS-empowered solutions. Yet, to demonstrate the practical feasibility of an advancing technology such RISs in ongoing next-generation network operations, test case validations, field tests, and real-life experiments are crucial. 
More recently, a set of experimental research based on developed testbeds has been conducted, and realistic transmission scenarios of the RISs have been investigated.
In this context, \cite{araghi2022reconfigurable} initially proposed a custom-developed continuous phase-shifted RIS operation in an indoor environment. Continuous phase surfaces serve more degrees of freedom than discrete phase shift RISs in terms of manipulating radio patterns owing to a more prosperous number of tunable phases. By employing this continuous phase surface, this study utilized software-defined radio (SDR) broadcasting video signals at the carrier frequency of 3.5 GHz. The researchers also investigated the received power performance for different receiver locations located at different angles with respect to the RIS in an L-shaped indoor environment \cite{araghi2022reconfigurable}. Furthermore, an RIS is also deployed in another similar L-shaped indoor test scenario for coverage enhancement purposes, where only a non-line-of-sight channel is available \cite{rains2021high}. This system deploys an RIS with several columns of sub-wavelength unit cells where each has a 3-bits phase resolution due to their custom hardware. This work also examined the received power performance of users positioned at various positions with monopole and horn antennas.
\begin{figure*}[t]
\centering
\begin{minipage}[h]{0.45\linewidth}
    \centering
    \includegraphics[width=\linewidth]{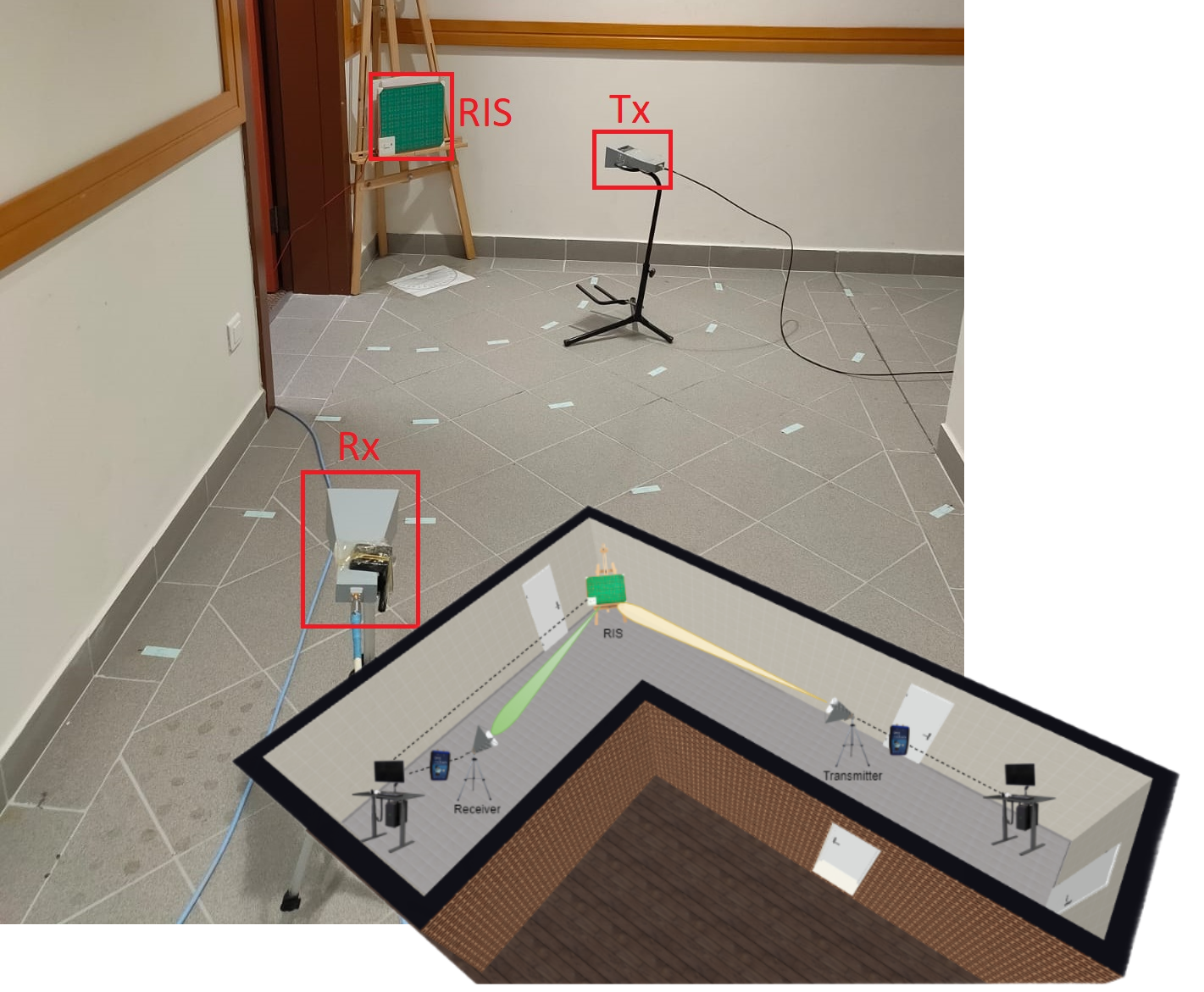}
    \subcaption{}
\end{minipage}
\begin{minipage}[h]{0.40\linewidth}
    \centering
    \includegraphics[width=\linewidth]{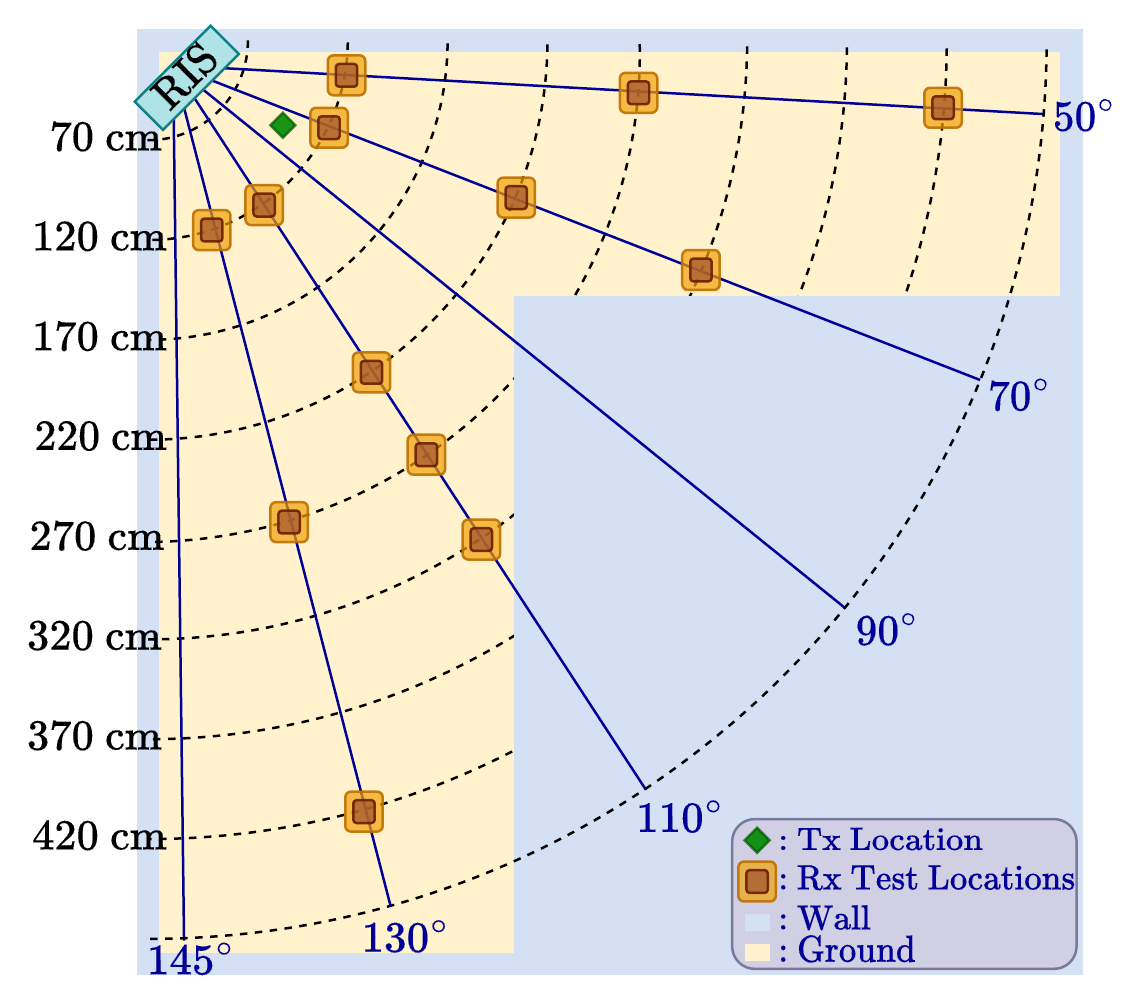}
    \subcaption{}
\end{minipage} \vspace{-0.2cm}
\caption{(a) The considered measurement setup of the RIS-aided wireless communication system in an indoor office environment, and (b) the top view of the setup with all Rx test locations.}
\label{fig:setup} \vspace{-0.5cm}
\end{figure*}

Another experimental scope is over 1-bit RISs for self-adaptive reflection, which is tested in indoor/outdoor environments in \cite{selfadaptive}. Furthermore, the performance of binary unit cells to control millimeter-wave (mm-Wave) beamforming in both near-field and far-field conditions is investigated in \cite{gros2021reconfigurable}. 1-bit RIS as a passive access point extender is proposed for mm-Wave communications in \cite{popov2021experimental}. Four-state reflection in sub 6GHz frequencies is studied in \cite{dai2020reconfigurable}. Moreover, instead of discrete phase shift designed schemes, \cite{fara2022prototype} implemented a continuous phase reflecting scheme driven by a varactor-controlled surface. Finally, \cite{zhang2022intelligent} investigated omni-scope transmission exploiting the joint progress of reflective and refractive features of RIS. 

Against this background, the literature is still premature for real-world experimental applications accommodating scenarios implying applicability. Proposed systems in the literature reveal some challenges that restrict their viability in practical use cases due to their computational burden and complexity. At this point, improving and optimizing RIS control processes becomes critical to obtain the optimal gains that can be provided by the existing RIS technologies.
Given that the first step toward RIS-assisted systems is passive beamforming techniques, the three targets of this work are (i) to realistically analyze the behavior of the RIS with respect to the receiver allocation points, (ii) to present a new line of practical applications by incorporating a codebook design into RIS without going through the optimization of RIS configurations, and (iii) to employ an efficient grouping scheme for the RIS elements in order to reduce the training time spent in the process of deciding the optimal RIS configurations. 

The organization of the paper is as follows; in Section II-A and Section II-B, system architecture, signal model, and measurement setup is given. The iterative approach to solving the RIS optimization problem is described in Section II-C. Section II-D reveals a number of measurement scenarios that lead to the investigation of the performance of the introduced RIS-assisted setting. Results corresponding to measurements are presented in Section III. Section IV concludes the paper.

\section{System Architecture}
In this section, we describe the technical details of our RIS-assisted transmission setup. Then, the RIS configuration and the proposed indoor transmission protocol are summarized by investigating efficient terminal positioning. \vspace{-0.1cm}

\subsection{System and Signal Model}
As illustrated in Fig. \ref{fig:setup}, we consider a basic wireless communication system where the RIS assists the signal transmission between a single-antenna transmitter (Tx) and receiver (Rx). The RIS utilized in the system model has a form of a uniform planar array (UPA) with $N=N_x\times N_y$ elements with binary phase shifts of $0^\circ$ and $180^\circ$. The channels from the Tx and Rx to the RIS are defined as $\mathbf{h}\in \mathbb{C}^{N}$ and $\mathbf{g}\in \mathbb{C}^{N}$, respectively. The direct line-of-sight (LoS) channel between the Tx and Rx is described as $h_{LoS}\in \mathbb{C}$. Accordingly, with $x[k]$ being the complex baseband transmitted signal, the received signal $r[k]$ can be written as
\begin{equation}
    r[k] = \left(\mathbf{g}^H\mathbf{\Theta} \mathbf{h}+h_{LoS} \right)x[k]+n[k]
\end{equation}
where $n[k]\sim \mathcal{N}_{\mathbb{C}}\left(0,\sigma_n^2\right)$ represents the noise component at the receiver, and $\mathbf{\Theta}$ represents the RIS configuration matrix, defined as $\text{diag}\left\{\alpha_{1} e^{j \phi_{1}}, \cdots, \alpha_{N} e^{j \phi_{N}}\right\}$, whose diagonal entries denote the states of the RIS elements. Since in our considered prototype, the RIS elements are passive and binary-phased, $\alpha_{n}$ is constant and $\phi_{n}$ takes values of only $0^\circ$ and $180^\circ$. Furthermore, the LoS component of the received signal is neglected since, as seen in Fig. \ref{fig:setup}, horn antennas, whose half beamwidth is $40^\circ$, are used in the Tx and Rx to transmit and receive signals, and the direct link between the terminals are out of the beamwidth of the antennas. 

One of the primary purposes of using an RIS in the transmission is to increase the signal-to-noise ratio (SNR) at the Rx by enabling anomalous reflections from the RIS, thereby enhancing the signal coverage. The SNR is proportional to end-to-end gain $\left|\mathbf{g}^H\mathbf{\Theta} \mathbf{h}\right|^2$, neglecting the LoS link between the terminals due to the non-intersecting antenna beams and can be adjusted by modifying the RIS configuration matrix $\mathbf{\Theta}$. Thus, the optimum configuration matrix of the RIS, $\mathbf{\Theta}^*$, which maximizes the end-to-end gain, is determined as the following expression:
\begin{equation}
\begin{aligned}
\mathbf{\Theta}^*=\arg &\max_{\phi_{n}, \forall  n} \left|\mathbf{g}^H\mathbf{\Theta} \mathbf{h}\right|^2 \\
&\text{ s.t. } \phi_{ n} \in\left\{0^{\circ}, 180^{\circ}\right\}, \quad \forall  n .
\end{aligned}\label{eq:max}
\end{equation}
In our practical setup, based on the maximization problem given in (\ref{eq:max}), without going through the channel estimation procedure, the objective function is considered as the average received signal power. The received signal powers are measured, and the objective function is optimized with an iterative method to obtain the RIS configuration, which maximizes the received signal power and focuses the reflected beam toward the Rx.

\subsection{Measurement Setup}
The considered system architecture of our RIS-assisted measurement setup is illustrated in Fig. \ref{fig:setup}. The measurement campaign is conducted in an L-shaped hallway of an indoor office environment with the help of two ADALM-PLUTO software-defined radio (SDR) modules for transmission and reception of over-the-air signals, $x[k]$. In the measurement setup, the RIS is mounted on the hallway's corner while the antennas of the Tx and Rx SDRs are positioned at various angles and distances from the center of the RIS throughout the measurement process. 

The RIS prototype developed by Greenerwave\cite{greenerwave} is used to focus the transmitted signal in the desired direction via software-controlled phase shifts. The working bandwidth of the RIS is $1$ GHz around the center frequency of $5.2$ GHz. The RIS has a UPA structure with a $10\times 8$ element grid, yet the right corner of the $2\times 2$ part is preserved for its controller, ending up with a total of $76$ reflecting elements whose reflection coefficient can be adjusted with PIN diodes. The PIN diodes can configure the phase shifts of the reflecting elements as $0^\circ$ or $180^\circ$. Furthermore, each element contains two PIN diodes corresponding to horizontal and vertical polarization, resulting in four possible states for each element. 

\subsection{Measurement Steps}
Through the measurement process, the Tx SDR sends a single-tone sinusoidal signal of $100$ kHz at the carrier frequency of $5.2$ GHz toward the RIS using a horn antenna, and the reflected signal from the RIS reaches the Rx SDR to measure the received signal power at $5.2$ GHz. The configuration for the RIS, which directs the reflected beam toward the Rx SDR, is determined by observing the received signal power. Since all possible RIS configurations of $4^{76}$ are not efficient to be applied through an exhaustive search algorithm, a straightforward iterative method \cite{popov2021experimental} attempts to find the RIS configuration, which will allow incoming signals to be reflected at the desired angle. Initially, all elements are set to OFF, indicating that the RIS is in no-phase-shift mode. Then, the four possible states of the first element are applied, and the RIS is kept in the state, resulting in the maximized received signal power at the Rx SDR. After determining the state of the first element, the same procedure is individually repeated for each of the remaining elements. The summary of the iterative phase adjustment method is given in Algorithm \ref{alg:ite}, where the measured received power and the obtained maximum received power are denoted as $P_r$ and $P_{max}$, respectively. 

\begin{algorithm}[t]
\vspace{0.1cm}
\caption{Iterative Method} \label{alg:ite}
\begin{algorithmic}[1]
\renewcommand{\algorithmicrequire}{\textbf{Input:}}
\renewcommand{\algorithmicensure}{\textbf{Output:}}
\REQUIRE $P_r$, number of RIS elements ($N$), and number of RIS states for one element ($P$)
\ENSURE  States
\STATE initiate States of RIS elements with OFF
\STATE $P_{max}$ $\gets 0$ 
\FOR {$i = 1$ to $N$}
\FOR {$j = 1$ to $P$}
\STATE apply state $j$ to $i$th element
\STATE measure $P_r$
\IF {($P_r>P_{max}$)}
\STATE update States
\STATE  $P_{max}$ $\gets$ $P_r$
\ENDIF
\ENDFOR
\ENDFOR
\RETURN States 
\end{algorithmic} \vspace{-0.1cm}
\end{algorithm}
In Algorithm \ref{alg:ite}, the received power is calculated by using the time samples captured at the center frequency of $5.2$ GHz and moved to the complex baseband by the Rx SDR. Since ADALM-PLUTO SDR has a 12-bit analog-to-digital converter, the sampled complex baseband signals can be represented as integers in the range of $(-2047, 2048]$. Therefore, the received signal power is computed in the decibels relative to full scale (dBFS) via the equation:
\begin{equation}
    P_{dBFS}=10\log_{10} \left(\frac{1}{K}\sum\nolimits_{k}^K |r[k]|^2\right)
\end{equation}
where $P_{dBFS}$ denotes the average power of the received complex baseband signal $r[k]$, and the total number of received samples is $K$. 

\subsection{Measurement Scenarios}
In the considered setup illustrated in Fig. \ref{fig:setup}, we aim to enhance the signal coverage in an indoor environment by increasing the received signal power via the deployment of an RIS to the communication network. In the transmission and reception of the signals, horn antennas provide control over the direction of the signals, while the considered measurement environment involving an L-shaped hallway prevents the direct link between the Tx and Rx SDR. 

The hallway is formed as a grid according to angles, defined clockwise from the RIS surface as $\{50^\circ,70^\circ,90^\circ,110^\circ,130^\circ,145^\circ \}$, and distances from the RIS as $\{70,120,170,220,270,320,420 \}$ cm. As shown in Fig. \ref{fig:setup}(b), the intersections of the lines of angles and distances denote our grid representation, which is used in the practical measurements. The Tx SDR is placed on a fixed point at $78^\circ$, and $100$ cm directed to the RIS, and the Rx SDR's antenna, which is considered to belong to a mobile user, is moved between the grid points based on the measurement scenarios.

Three different experiment scenarios are investigated in the course of our measurement campaign, which aims to focus the reflected beam of the RIS toward the user: In the first trial, the Rx SDR is positioned at different angles and distances to observe the performance of the RIS for the considered test points. In the second experiment, a codebook is formed for pre-determined test points on the grids to eliminate the feedback channel from the Rx to the RIS, which is employed during the process of determining the RIS configuration on the move. Throughout the experiment, the Rx SDR's antenna is moved along the test points at the interior regions of the grids. The performance comparisons for the test points are conducted between the offline codewords from the saved codebooks and the RIS configurations of the online iterative method, which is obtained from Algorithm \ref{alg:ite}. Finally, in the last scenario, the effect of grouping the RIS elements for joint phase adjustment is evaluated by reducing the training time required for selecting the configuration of the RIS. 

\section{Measurement Results}
In this section, we present results from several experiments using the RIS prototype to test the performance of the RIS in real-world communication scenarios. Moreover, an efficient grouping strategy for the RIS elements is investigated in the L-shaped indoor environment to decrease the high training overhead burden on the RIS controller. 

\subsection{Performance Analysis for Various Angles and Distances}
\begin{figure}[t]
    \centering \vspace{0.1cm}
    \includegraphics[width=0.75\linewidth]{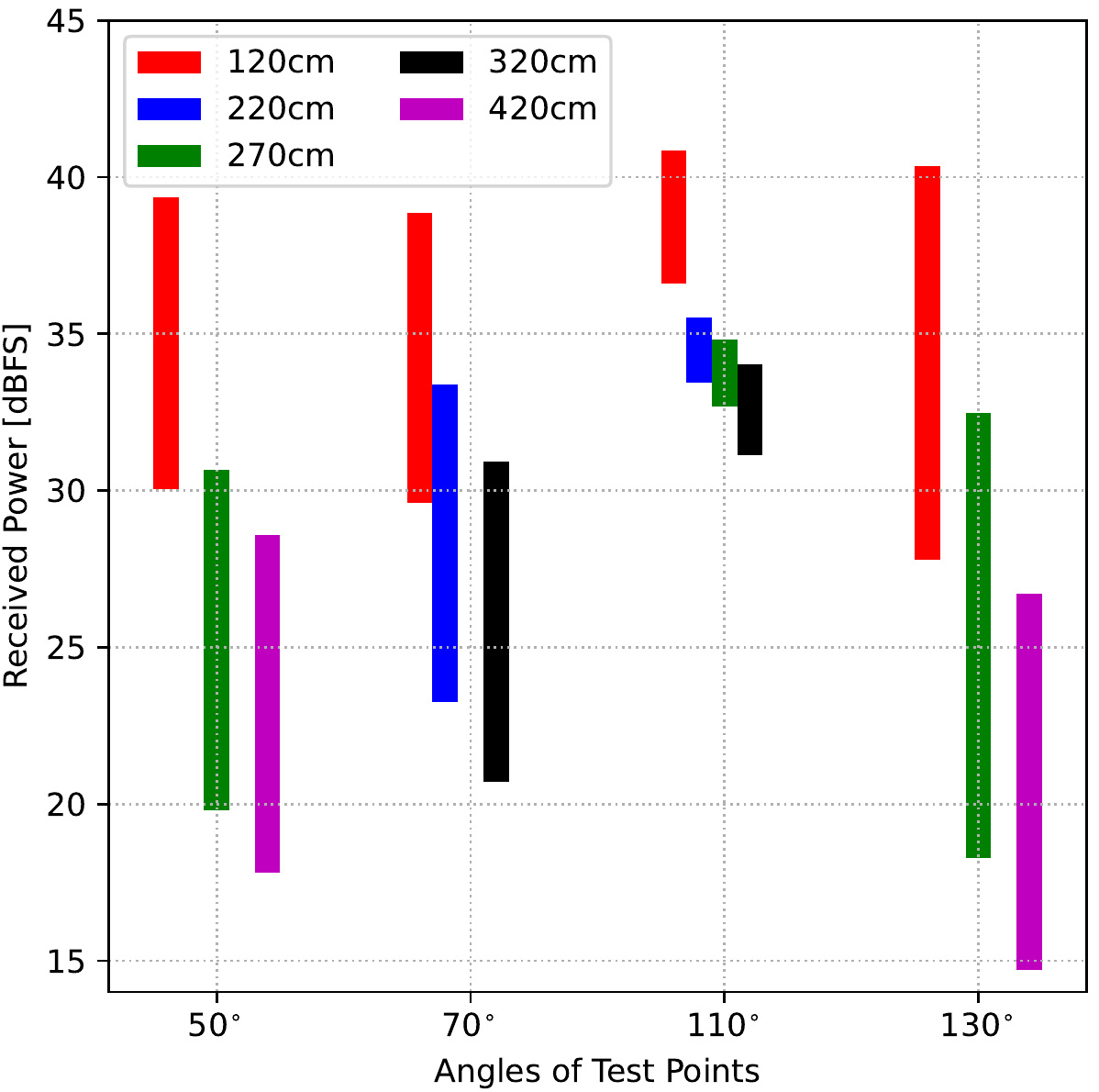}
    \caption{The received power comparison for various locations of the user.}
    \label{fig:grids} \vspace{-0.6cm}
\end{figure}

Throughout the experiments, the RIS is controlled with a Python script, which takes a Boolean array of $152$ elements corresponding to PIN diodes of the horizontal and vertical polarization states. In the Boolean array, $0$ and $1$ correspond to no-phase-shift and $180^\circ$-phase-shift mode, respectively. The sampling rate of the Tx and Rx SDR are set to $1$ MHz. The buffer size of the Rx SDR, the number of time samples to calculate the received signal power, is selected as $10000$.

The improvement in the power of the received signal via the RIS at the Rx will largely depend on the distance between the terminals. On the other hand, the variations in the position of Rx will alter the angle-of-arrival of the incoming signal from the RIS as well as the RIS-Rx distance. In order to analyze the performance of the RIS-assisted transmission systems for varying angles, the Rx SDR's antenna is placed at 13 different test points on the grid, which are illustrated in Fig. \ref{fig:setup}(b) as brown squares. 

Fig. \ref{fig:grids} shows the comparison of the received signal power for these test points when the RIS configuration obtained with Algorithm \ref{alg:ite} is applied. For example, the green bar for the angle of $50^\circ$ in Fig. \ref{fig:grids} illustrates that the received signal power is around 20 dBFS at the beginning of the iterations, and it increases by 11 dB and reaches approximately 31 dBFS at the end of the iterations. As clearly seen in Fig. \ref{fig:grids}, the RIS performances at different distances for the same angle exhibit very similar patterns of enhancements on the received signal power. In fact, the increase in the received signal power is around 10 dB for the angles of $50^\circ$ and $70^\circ$ and approximately 12 dB for the angles of $130^\circ$. For the case of $110^\circ$, the RIS can only boost the received signal power by around 4 dB since the Rx SDR has higher received signal power because of the symmetrical position of the Rx with respect to the Tx. 

\subsection{Codebook Implementation for Mobile Receiver}
\begin{figure}[t]
\centering
\vspace{0.15cm}
\begin{minipage}[h]{0.32\linewidth}
    \centering
    \includegraphics[width=\linewidth]{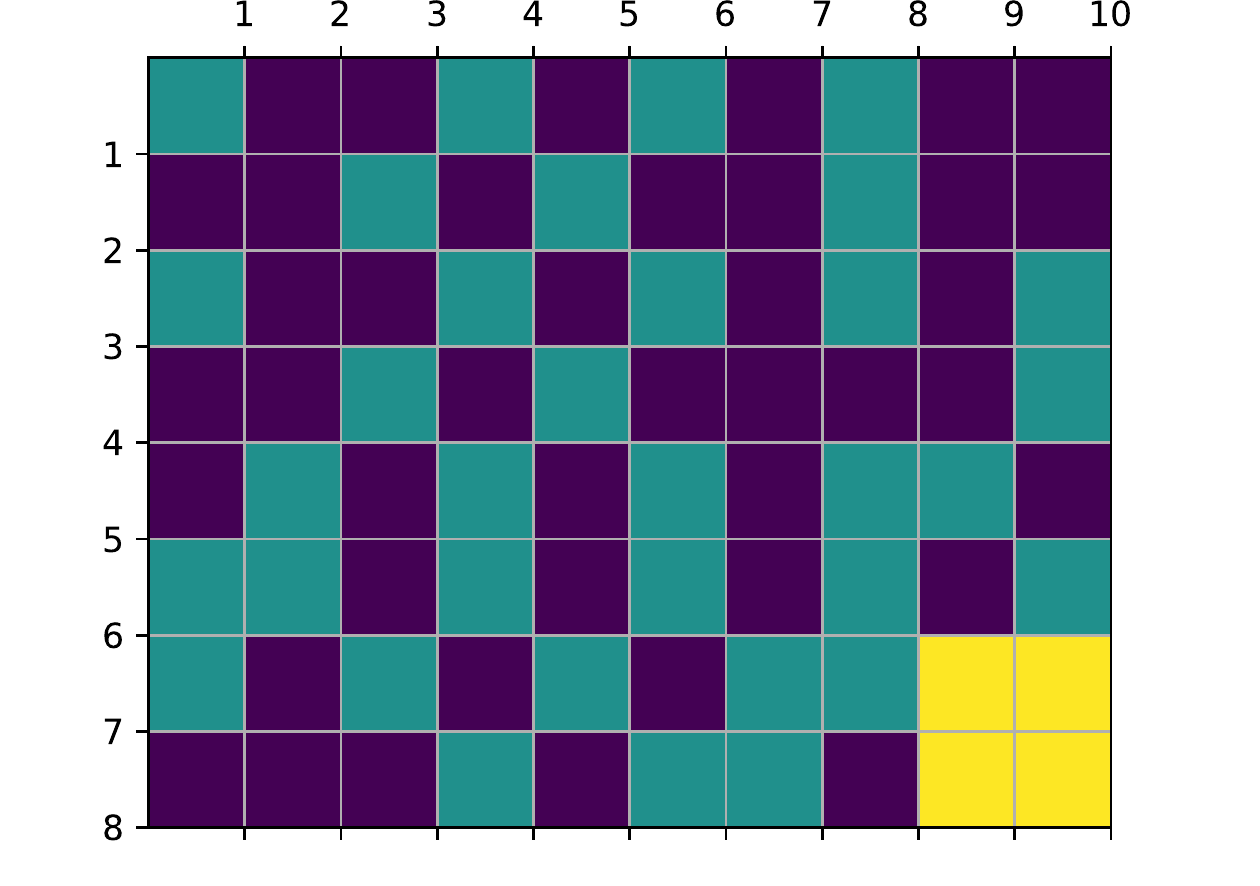}
    \subcaption{}
\end{minipage}
\begin{minipage}[h]{0.32\linewidth}
    \centering
    \includegraphics[width=\linewidth]{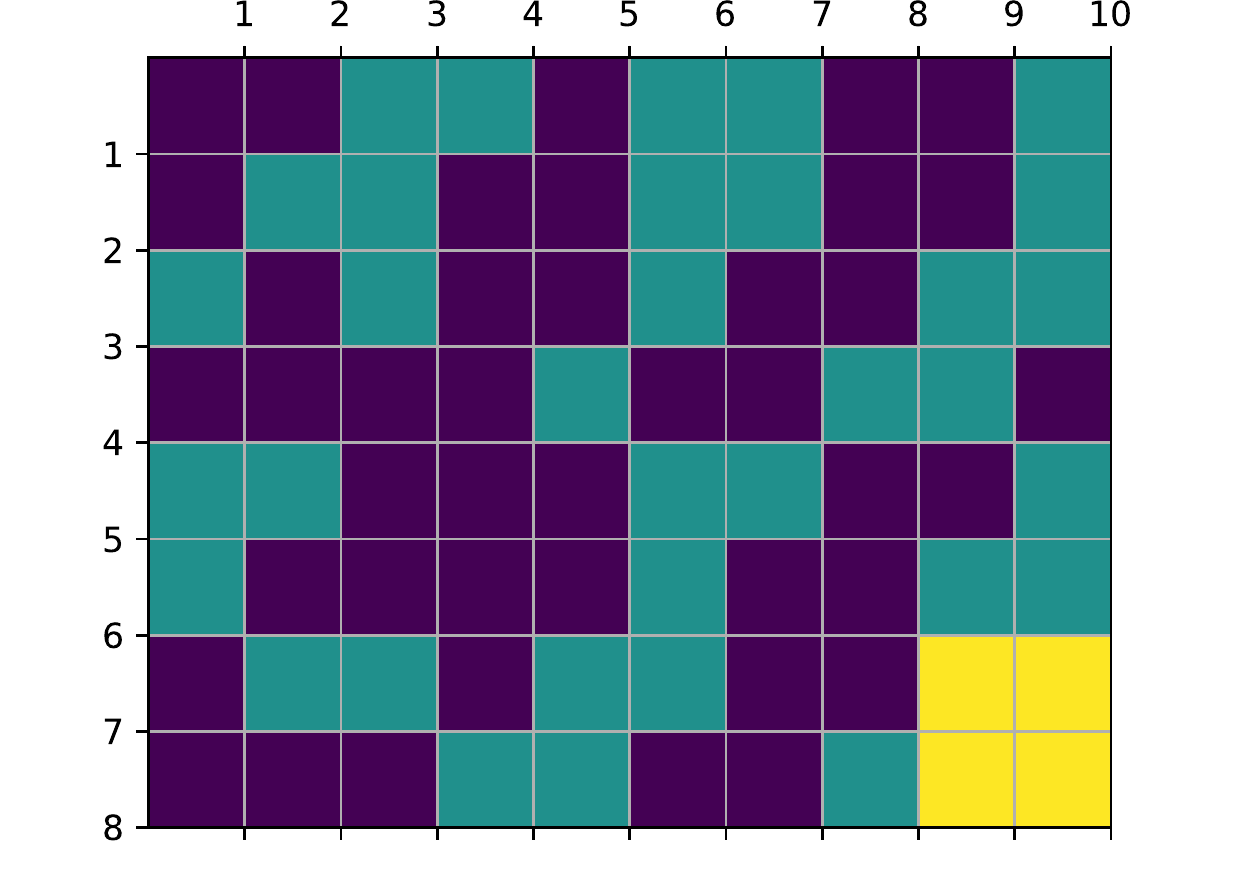}
    \subcaption{}
\end{minipage}
\begin{minipage}[h]{0.32\linewidth}
    \centering
    \includegraphics[width=\linewidth]{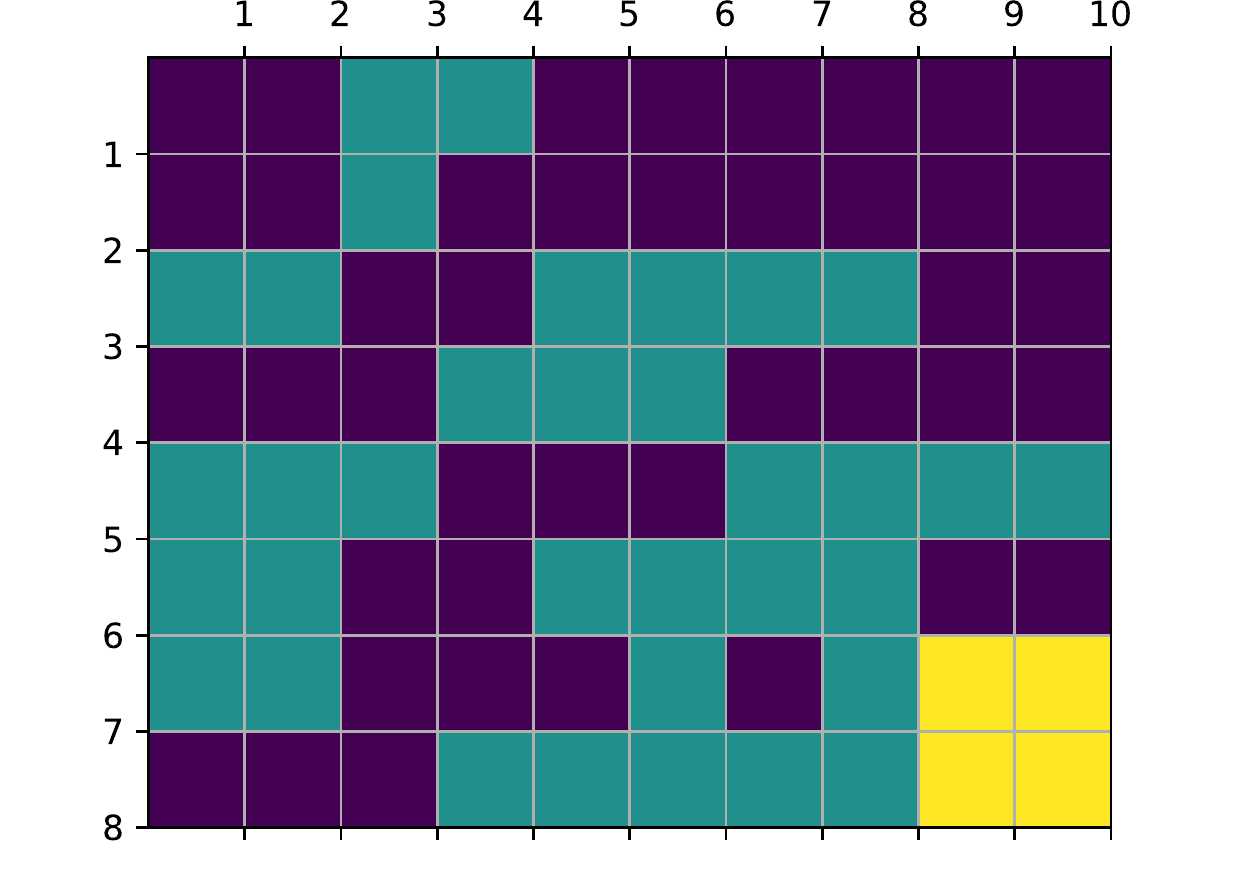}
    \subcaption{}
\end{minipage}
\begin{minipage}[h]{0.32\linewidth}
    \centering
    \includegraphics[width=\linewidth]{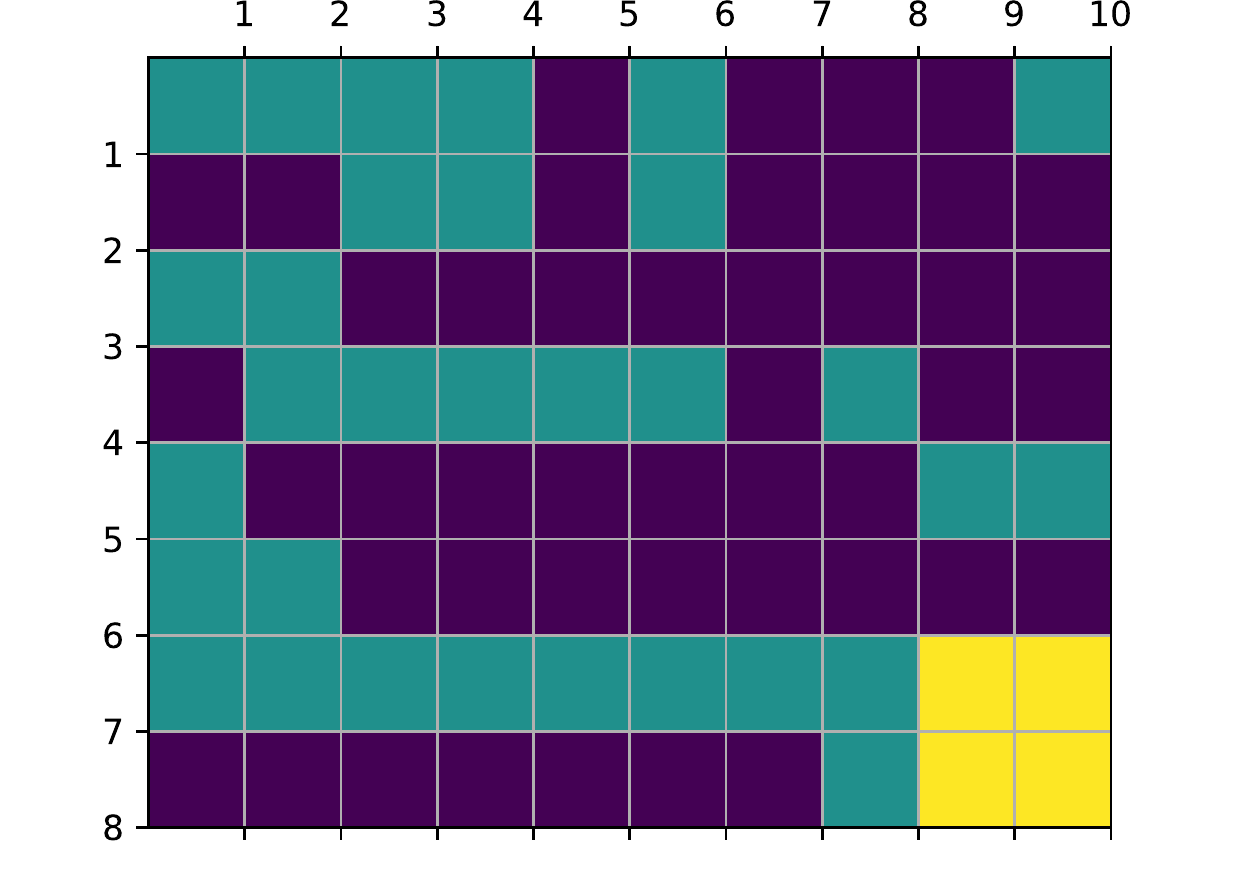}
    \subcaption{}
\end{minipage}
\begin{minipage}[h]{0.32\linewidth}
    \centering
    \includegraphics[width=\linewidth]{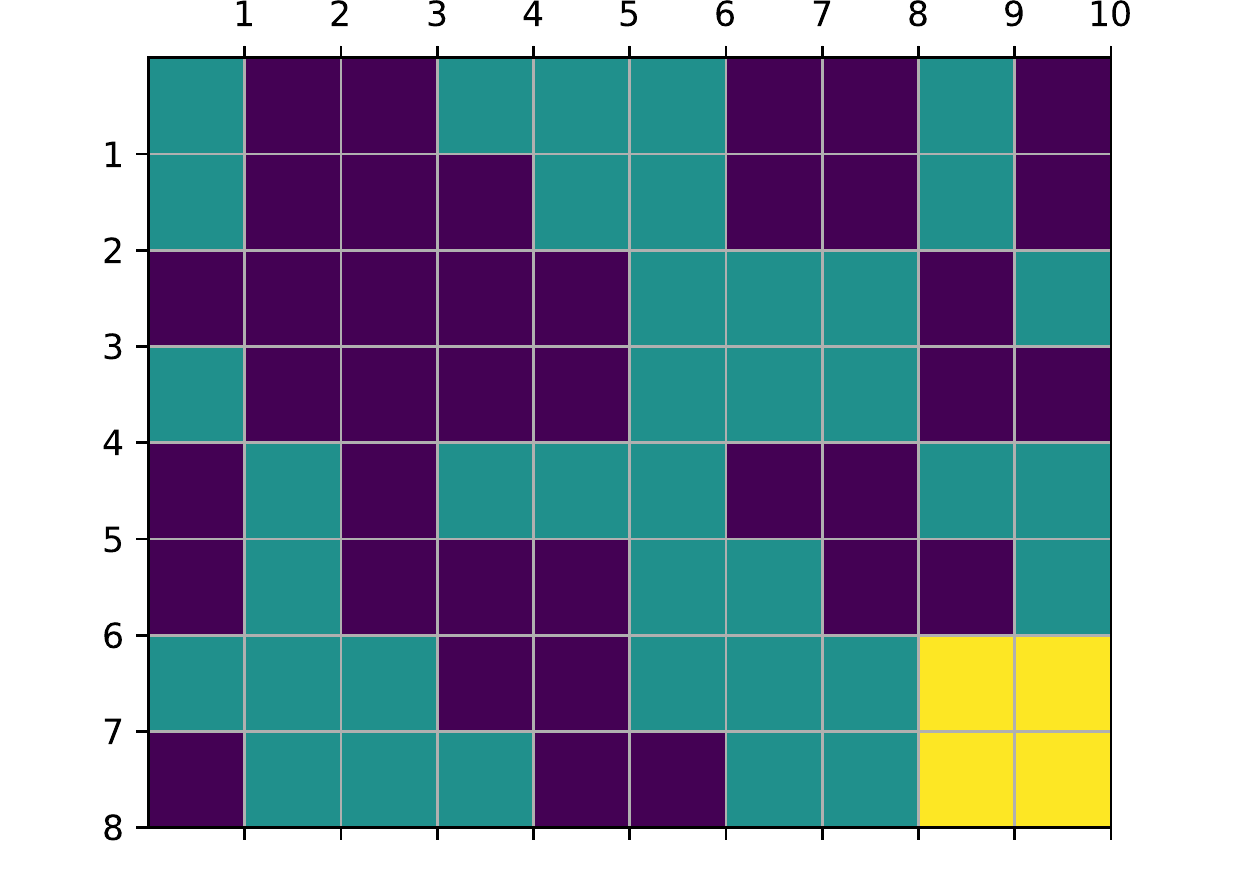}
    \subcaption{}
\end{minipage}
\begin{minipage}[h]{0.32\linewidth}
    \centering
    \includegraphics[width=\linewidth]{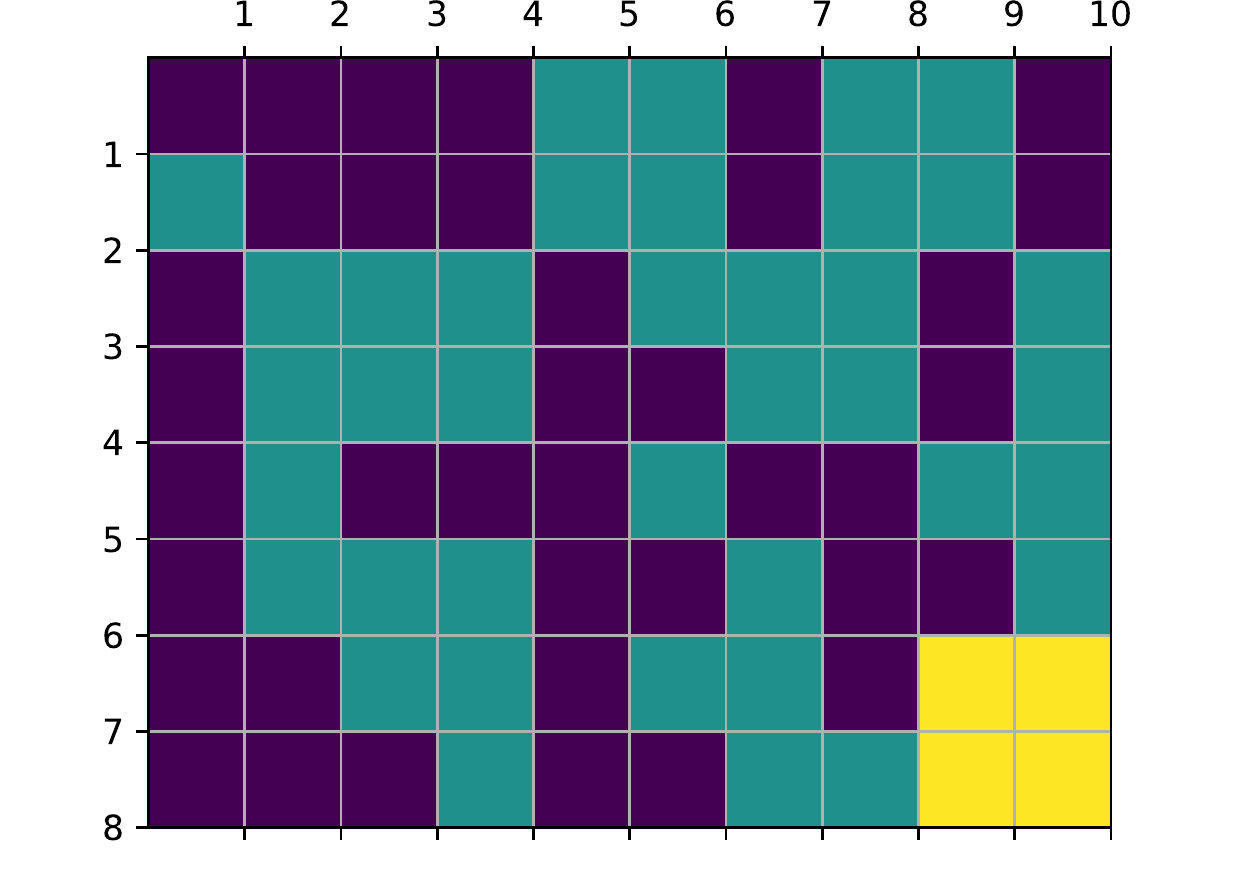}
    \subcaption{}
\end{minipage}
\caption[]{The codewords of the horizontal PIN diodes on/off configurations at the distance of $170$ cm, and the angles of (a) $50^\circ$, (b) $70^\circ$, (c) $90^\circ$, (d) $110^\circ$, (e) $130^\circ$, and (f) $145^\circ$.}
\label{fig:codewords} \vspace{-0.3cm}
\end{figure}

For mobile users, the RIS configuration has to be adjusted on the move in the coherence time of the channel between the RIS and Rx. The time spent during an iterative method like Algorithm \ref{alg:ite} can exceed this, which the channel can be considered as stationary. Furthermore, for increasing data rates with reduced symbol duration, the time overhead of Algorithm \ref{alg:ite} cannot be ignored. Thus, the implementation of a codebook, which contains the recorded RIS configurations for pre-determined points, can satisfy the time limitation of the channel by applying the saved RIS configurations based on the given locations of the terminals. Furthermore, the employment of a codebook for the RIS configurations eliminates the dependency of the feedback channel between the Rx and RIS, which is utilized to select the RIS configuration. Namely, the codebook-based approach can be employed by only requiring the geometric positions or the departure/arrival angles of the terminals. With this consideration, a codebook containing codewords of the RIS configurations for all the grid angles at $170$ cm is generated using Algorithm \ref{alg:ite} to be used afterward. The formed codewords of the horizontal PIN diodes on/off configurations are given in Fig. \ref{fig:codewords}. In the given configuration instances in Fig. \ref{fig:codewords}, the purple color represents the no-phase-shift mode of the RIS while the green color indicates the $180^\circ$-phase-shift mode and the yellow part corresponds to the RIS controller, which has no reflecting elements.

\begin{figure}[t]
    \centering
    \includegraphics[width=0.7\linewidth]{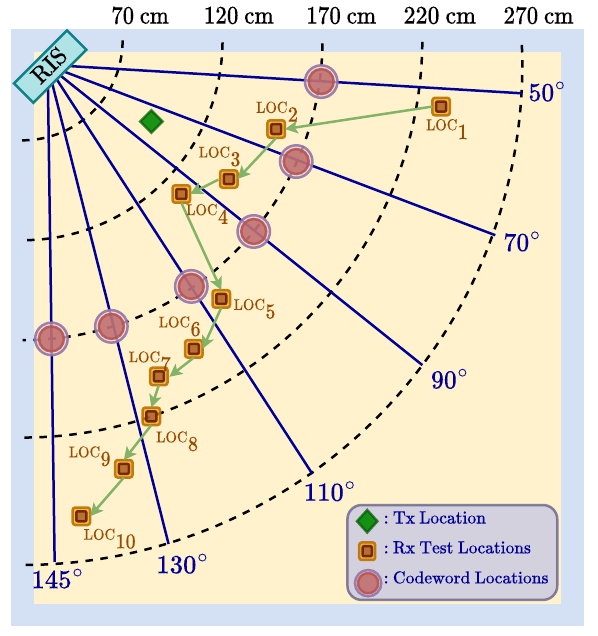}
    \caption{Top view of the considered RIS-assisted transmission scenario under the varying locations of the user.}
    \label{fig:moving} \vspace{-0.6cm}
\end{figure}

The Rx SDR's antenna is moved along the test points in the interior regions of the grids, and Fig. \ref{fig:moving} illustrates its path along the hallway. In Fig. \ref{fig:moving}, pink circles represent the points where RIS configurations that maximize their received power are recorded to generate the codebook, while brown squares denote the locations of points where the mobile Rx moves.
\begin{figure}[t]
    \centering 
    \includegraphics[width=0.85\linewidth]{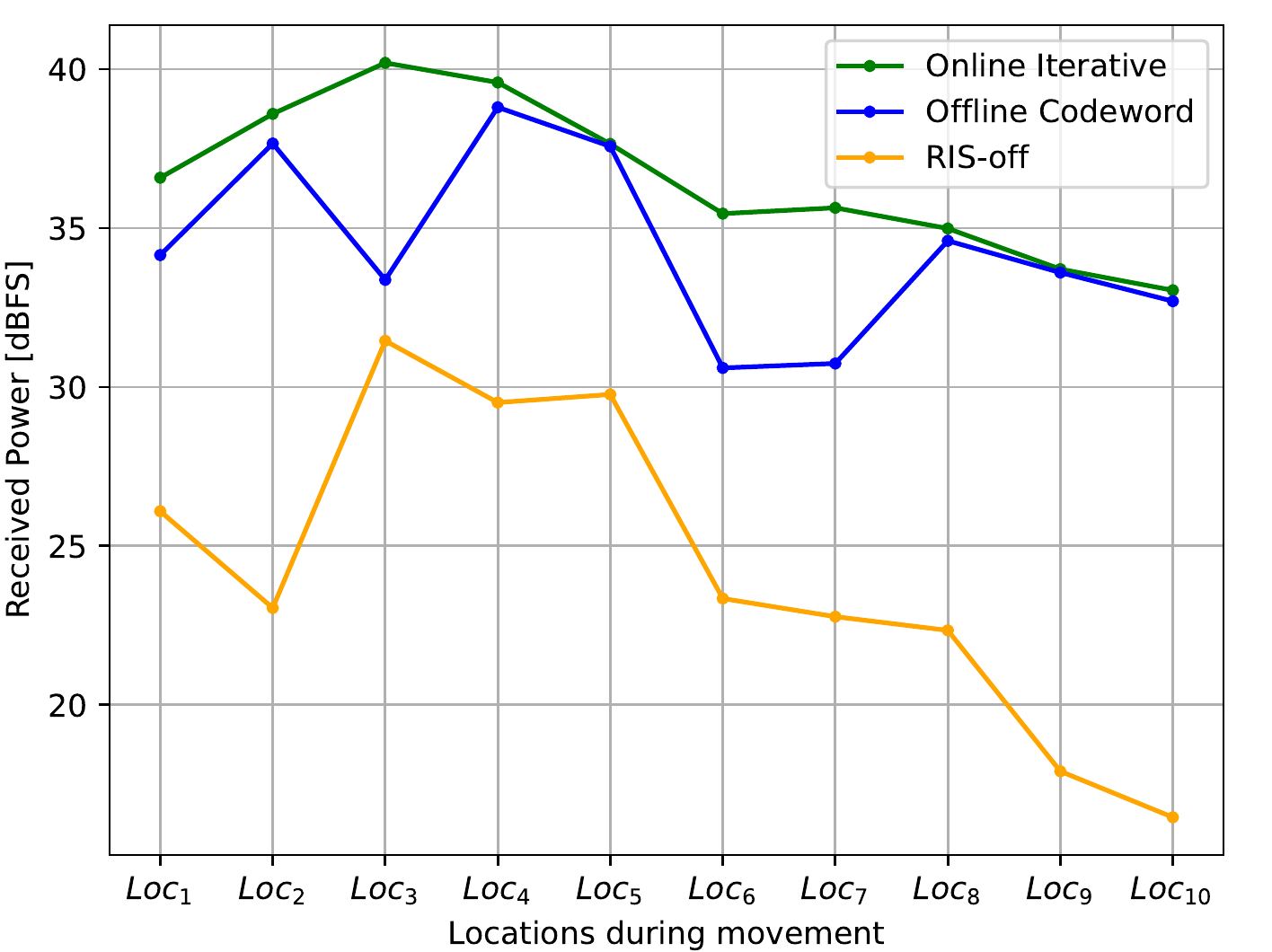}
    \caption{The comparison of received signal power during\\ the movement of the Rx.}
    \label{fig:codebook} \vspace{-0.4cm}
\end{figure}

During the movement of the Rx SDR's antenna, the corresponding codewords corresponding to the closest reference points are utilized to improve the performance of the Rx. The received signal powers are measured and compared for the three cases where: (i) the RIS is in no-phase-shift mode, (ii) the offline codeword of the stored codebook of the closest point is applied to the RIS, and (iii) the configuration of the online iterative method, which is derived via Algorithm \ref{alg:ite} on the test point, is set to the RIS. Fig. \ref{fig:codebook} shows the variation of the received signal powers of the mobile Rx through the test points for these three cases.

It can be observed from Fig. \ref{fig:codebook} that the RIS configured with the codewords tends to keep the received signal powers above 30 dBFS even when the Rx moves further distances. Moreover, the performance of our codebook for locations where its angles are close to the angles of the saved codewords is observed to be very close to the results of online iterative configurations. Hence, using the offline codebook approach, the RIS can achieve comparable performance, which is approximately 3 dB lower than the real-time iterative method, without going through repeated iterations of Algorithm \ref{alg:ite}. Accordingly, this approach can fulfill the time constraint of the channel with an aid of direction finding algorithm since switching a codeword from another in the RIS takes approximately 1 ms \cite{gros2021reconfigurable}.

\subsection{Effect of Grouping the RIS Elements}
\begin{figure}[t]
\centering
\begin{minipage}[h]{0.41\linewidth}
    \centering
    \includegraphics[width=\linewidth]{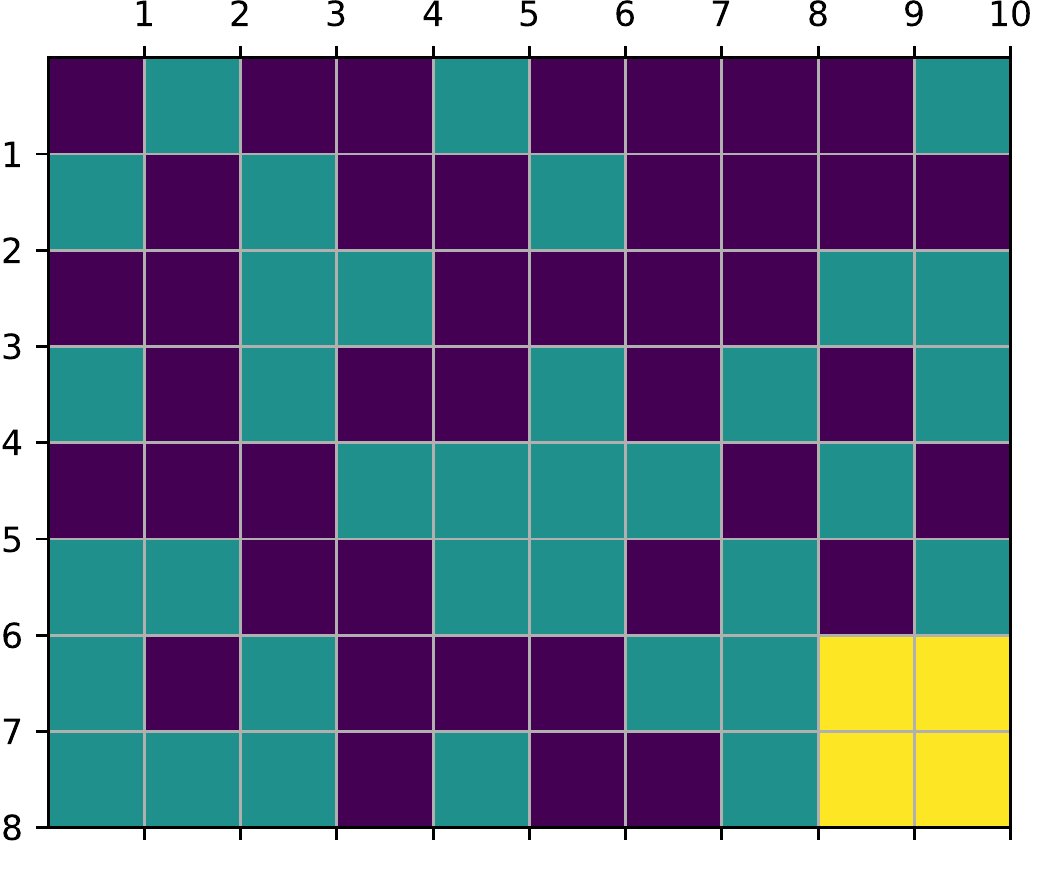}\vspace{-0.2cm}
    \subcaption{}
\end{minipage}
\begin{minipage}[h]{0.41\linewidth}
    \centering
    \includegraphics[width=\linewidth]{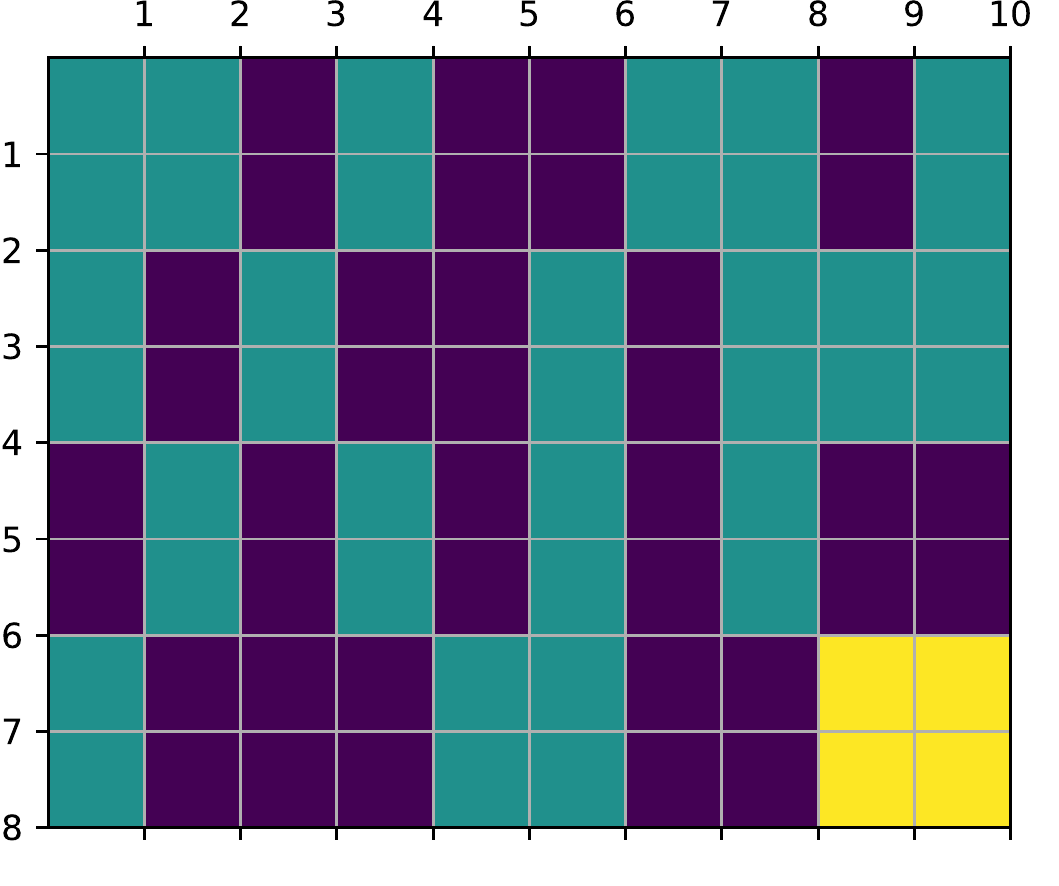}\vspace{-0.2cm}
    \subcaption{}
\end{minipage}
\begin{minipage}[h]{0.41\linewidth}
    \centering
    \includegraphics[width=\linewidth]{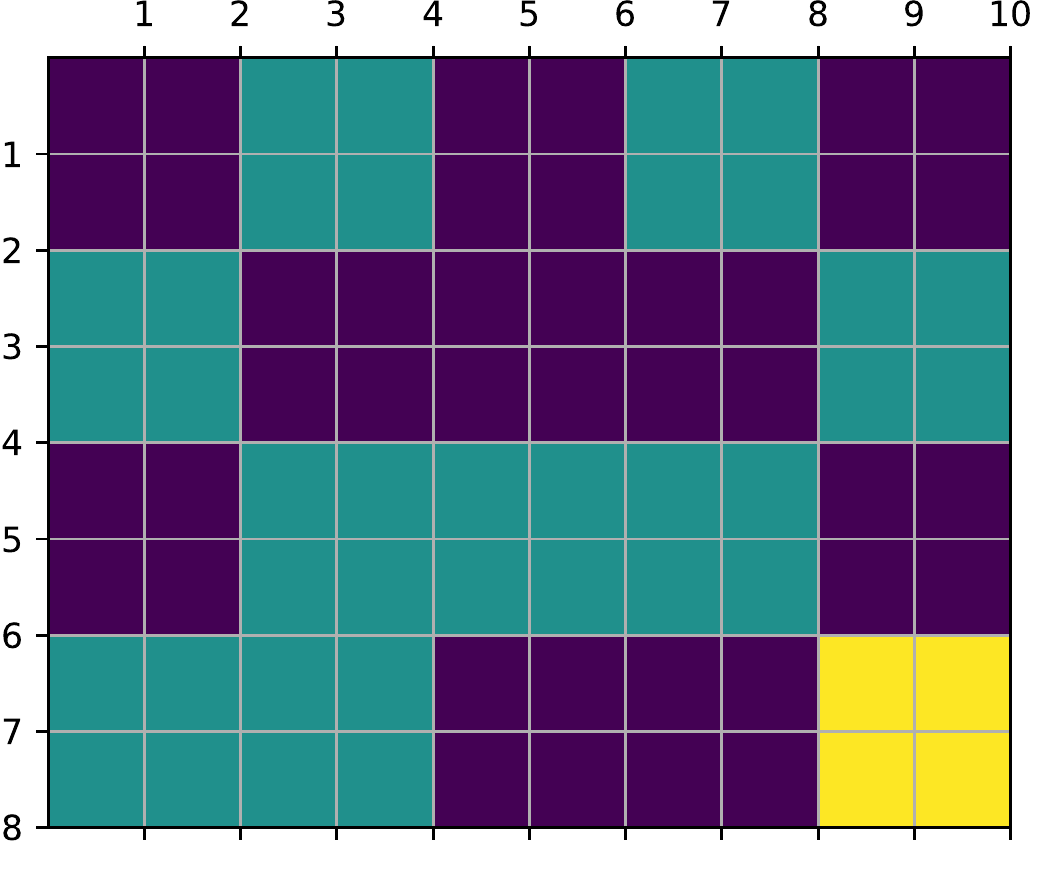}\vspace{-0.2cm}
    \subcaption{}
\end{minipage}
\begin{minipage}[h]{0.41\linewidth}
    \centering
    \includegraphics[width=\linewidth]{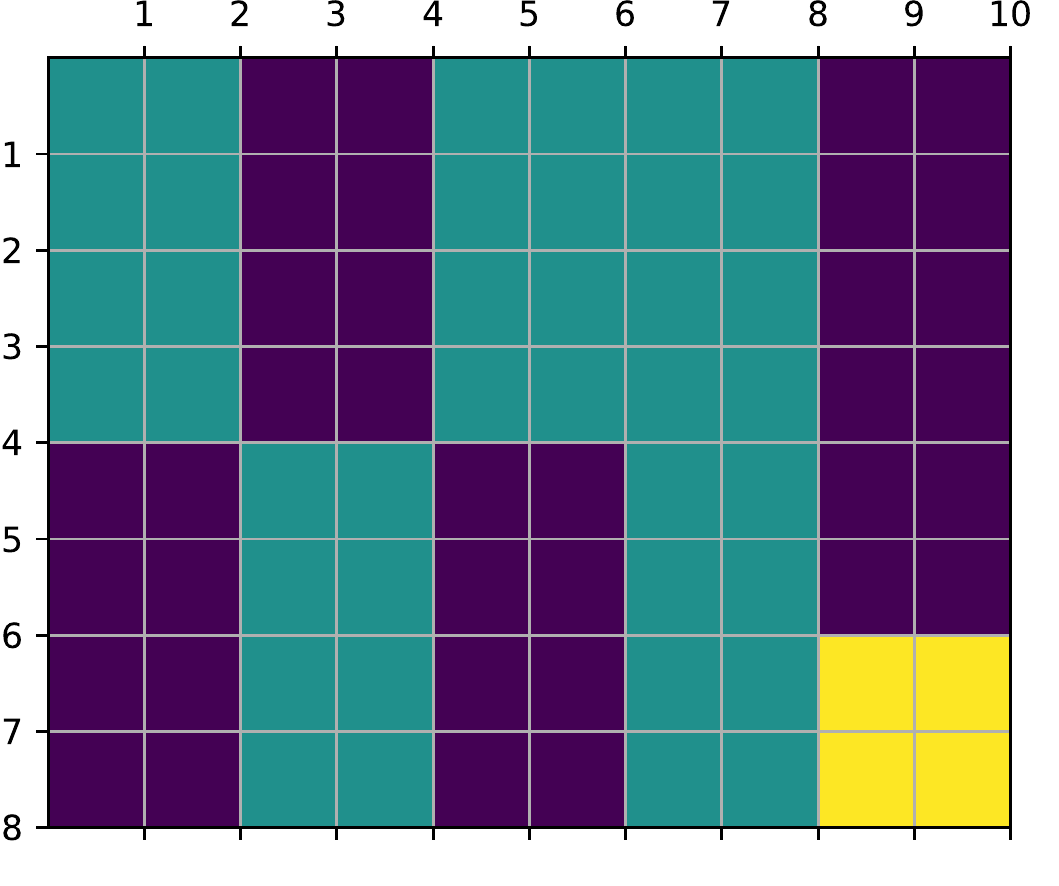}\vspace{-0.2cm}
    \subcaption{}
\end{minipage}
\caption[]{Examples the RIS configurations of (a) one-by-one, (b) two-by-two, (c) four-by-four, and (d) eight-by-eight grouping.}
\label{fig:groups} \vspace{-0.5cm}
\end{figure}
\begin{figure*}[t]
    \centering
    \includegraphics[width=0.9\linewidth]{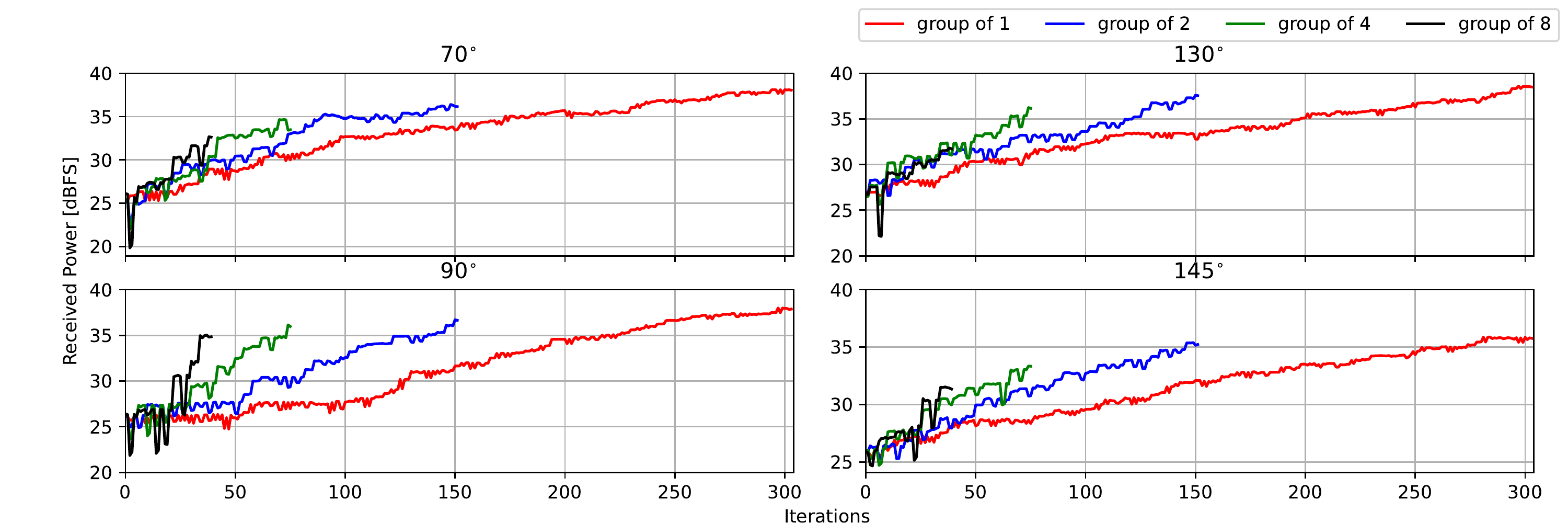}
    \caption{The received signal power during the iterations of Algorithm \ref{alg:ite} for the grouping approach.}
    \label{fig:grouping} \vspace{-0.3cm}
\end{figure*}

The number of RIS elements directly affects the training time required to determine optimal RIS configurations to steer the beam toward the Rx. The increasing number of RIS elements will also increase the reconfiguration overhead of the RIS. Therefore, one of the useful strategies for finding optimum RIS configuration is the sub-grouping of the RIS elements in the phase adjustment process. Another objective of grouping the RIS elements is to reduce the amount of time spent during the training process to generate the configuration that focuses the beam toward the Rx SDR. Fig. \ref{fig:groups} shows an example of the sub-grouped RIS configurations (only for one polarization of the RIS) with groups of two, four, and eight elements. The instances demonstrate how the grouping is applied.

The received signal powers for the Rx angles of $70^\circ$, $90^\circ$, $130^\circ$, and $145^\circ$ through the iterations of Algorithm \ref{alg:ite}, which are obtained from grouped elements, are shown in Fig. \ref{fig:grouping}. The given plots demonstrate the impact of grouping with two, four, and eight elements with respect to the increasing iteration. 

Fig. \ref{fig:grouping} illustrates that the groups of two and four elements mostly perform close to the one-by-one adjustment case. For instance, even at $90^\circ$, the group of eight elements increases the received signal power by around 3 dB less than the single-element iterations. For most of the angles, the groups of two and four elements accomplish very close received signal power, within a few dB, in comparison to the one-by-one case. Thus, employing the grouping approach would significantly reduce the time and effort required to optimize the RIS configurations. 

\section{Conclusion}
In this study, a real-world RIS has been employed in a system operating at $5.2$ GHz to manipulate the over-the-air signals in an indoor environment in order to enhance the signal coverage by improving the received signal power. With the help of the proposed iterative algorithm, the fully passive RIS provides a significant increase in the received signal power for different angles and distances from the RIS. In order to reduce the RIS configuration overhead, a phase shift codebook for the geometric positions of the Tx and Rx has also been designed to be implemented at the nearest test points. For our considered measurement setup, the saved codewords achieve similar performance in comparison to the online iterative results, and the RIS with the codewords is able to provide reliable transmission throughout the movement of the Rx. Finally, the proposed grouping strategy for the RIS elements is considered to reduce the training time needed to determine the RIS configurations. By means of our efficient grouping strategy, the training time is significantly decreased by losing only a few dBs. We conclude that our initial measurement campaign might be useful to more clearly understand the practical application aspects of RISs in challenging environments. 

\section*{Acknowledgment}
The authors would like to thank the Greenerwave team for their technical support with the RIS prototype. The statements made herein are solely the responsibility of the authors. We thank to StorAIge project that has received funding from the KDT Joint Undertaking (JU) under Grant Agreement No. 101007321. The JU receives support from the European Union’s Horizon 2020 research and innovation programme in France, Belgium, Czech Republic, Germany, Italy, Sweden, Switzerland, Türkiye, and National Authority TÜBİTAK with project ID 121N350. The work of E. Basar and Y. Gevez is also supported by TÜBİTAK under grant 120E401.
\vspace{-0.1cm}
\balance
\bibliographystyle{IEEEtran}
\bibliography{main.bib}
\end{document}